\documentclass[journal=jacsat,manuscript=article]{achemso}

\usepackage[version=3]{mhchem} 
\usepackage{xcolor}

\author{Frederico B. Sousa}
\affiliation{Departamento de F\'isica, Universidade Federal de Minas Gerais, Belo Horizonte, Minas Gerais 30123-970, Brazil}

\author{Lucas Lafeta}
\affiliation{Departamento de F\'isica, Universidade Federal de Minas Gerais, Belo Horizonte, Minas Gerais 30123-970, Brazil}

\author{Alisson R. Cadore}
\affiliation{MackGraphe – Graphene and Nanomaterials Research Center, Mackenzie Presbyterian University, Sao Paulo, 01302-907, Brazil}

\author{Prasana K. Sahoo}
\affiliation{Materials Science Centre, Indian Institute of Technology Kharagpur, Kharagpur, West Bengal, 721302, India}

\author{Leandro M. Malard}
\affiliation{Departamento de F\'isica, Universidade Federal de Minas Gerais, Belo Horizonte, Minas Gerais 30123-970, Brazil}
\email{lmalard@fisica.ufmg.br}

\title[An \textsf{achemso} demo]
  {Revealing  Interfaces of Two-Dimensional Lateral Heterostructures by Second Harmonic Generation}
\abbreviations{IR,NMR,UV}

\begin{document}

\begin{abstract}

The interface between two different semiconductors is crucial in determining the electronic properties at the heterojunction, therefore novel techniques that can probe these regions are of particular interest. Recently it has been shown that heterojunctions of two-dimensional transition metal dichalcogenides have sharp and epitaxial interfaces that can be used to the next generation of flexible and on chip optoelectronic devices. Here, we show that second harmonic generation (SHG) can be used as an optical tool to reveal these atomically sharp interfaces in different lateral heterostructures. We observed an enhancement of the SH intensity at the heterojunctions, and showed that is due to a coherent superposition of the SH emission from each material. This constructive interference pattern reveals a phase difference arising from the distinct second-order susceptibilities of both materials at the interface. Our results demonstrate that SHG microscopy is a sensitive characterization technique to unveil nanometric features in layered materials and their heterostructures.

\end{abstract}


 \noindent Vertical and lateral heterostructures based on two-dimensional (2D) materials have shown to provide a plethora of new possibilities by combining different layered crystals \cite{Geim2013, Novoselov2016, Liu2016}. New electronic states are formed at the interfaces (both vertical and lateral) of the materials and are of great importance to applications \cite{Wang2012} and to understand and tune the physical properties of these novel heterostructures \cite{Geim2013, Novoselov2016, Liu2016, Tartakovskii2020, Shree2021}. Lateral heterostructures (LHs) based on 2D transition metal dichalcogenides (TMDs) have shown a promising potential to be used as a p-n junction building blocks \cite{Li2016, Sahoo2019} in optoelectronic devices such as transistors \cite{Sarkar2015,Lin2015}, photodetectors \cite{Yu2013,Withers2015}, light-emitting diodes \cite{Withers2015} and photovoltaic cells \cite{Pospischil2014,Lee2014}. The electronic properties as bandgap, work function and spin-orbit coupling strength considerably vary between different TMD monolayers (1L) \cite{Wang2018}. Thereby, the engineering of these properties between different 1L-TMD leads to a range of distinct LHs possibilities. For instance, MoS$_{2}-$MoSe$_{2}$ \cite{Duan2014}, MoS$_{2}-$WS$_{2}$ \cite{Gong2014,Chen2015,Zhang2015,Sahoo2018}, MoS$_{2}-$WSe$_{2}$ \cite{Li2015}, MoSe$_{2}-$WS$_{2}$ \cite{Zhang2017}, MoSe$_{2}-$WSe$_{2}$ \cite{Huang2014,Gong2015,Zhang2015,Sahoo2018,Jia2020} and WS$_{2}-$WSe$_{2}$ \cite{Duan2014,Zhang2017} are types of 2D TMD LHs already synthesized by different growth methods. The heterojunctions of these LHs, i.e., the interfaces between both materials, have been widely investigated by different techniques such as scanning transmission electron microscopy (STEM) \cite{Duan2014, Gong2014, Chen2015, Zhang2015, Sahoo2018, Li2015, Zhang2017, Huang2014, Gong2015, Jia2020} and scanning probe microscopy (SPM) \cite{Chen2015, Sahoo2019(2)}, as well as Raman and photoluminescence (PL) spectroscopy \cite{Duan2014, Gong2014, Chen2015, Zhang2015, Sahoo2018, Li2015, Zhang2017, Huang2014, Gong2015, Jia2020}. Nevertheless, nonlinear spectroscopy in such nanostructures has been barely explored \cite{Zhang2015, Li2015}.

 Second harmonic generation (SHG) has been used to investigate a variety of condensed matter systems, ranging from bulk- to nanomaterials \cite{Butet2015, Blake2020}. In particular, SHG has also been used to probe interfaces in materials that are centrosymmetric due to the symmetry breaking at the surfaces and interfaces \cite{Heinz1982, Shen1989, Shen1989_2}, to provide information about the crystallography orientations of 2D materials \cite{Li2013, Malard2013, Kumar2013, Wang2019}, to probe grain boundaries defects \cite{Yin2014, bruno2020}, to identify modifications in the nonlinear optical susceptibility of layered materials \cite{Cunha2020}, and to access vertical heterostructures \cite{Hsu2014, Kim2020}. Since SHG is a coherent effect, interference phenomena play a major role on the efficiency of the light conversion \cite{Boyd2003}, and thus can be used to reveal otherwise hidden features due to the optical diffraction limit \cite{Yin2014, bruno2020}. Here, we systematically show a surprising and strong second harmonic (SH) emission that arises from the atomically sharp interfaces at the 2D TMDs LHs. We show that there is a constructive SH interference at these interfaces due to a coherent superposition of the SH emission from each material. Our results reveal a finite phase difference between both materials that accounts for their distinct second-order susceptibilities ($\chi^{(2)}$). Such result agrees with the measured SH energy dependence and polarization for different combination of TMDs and number of layers. To our knowledge, this is the first report of this effect and opens the opportunity to use 2D LHs as building blocks for photonics applications where the SH efficiency can be tuned by the appropriate choice of material, pump wavelength and polarization.

 The development of the one-pot growth of 2D TMD LHs via sequential edge-epitaxy by Sahoo \textit{et al.} \cite{Sahoo2018} enabled the high-quality and controllable fabrication of these heterostructures, features that were not simultaneously achieved by single-step \cite{Duan2014, Gong2014, Huang2014}, two-step \cite{Gong2015, Li2015} and multi-step \cite{Zhang2017} growth methods. The LHs grown by the one-pot method presented pure TMD domains as well as defect-free and sharp heterojunctions \cite{Sahoo2018, Sahoo2019, Xue2018, Berweger2020}. Nonetheless, these LHs has not been probed by SHG so far. The domains in our LHs samples were characterized by Raman spectroscopy as shown in Supporting Information (SI) Section S1. The details of sample preparation, SHG and Raman spectroscopy experiments are presented in Methods Section.

\begin{figure}[!htb]
 \centering
 \includegraphics[scale=0.5]{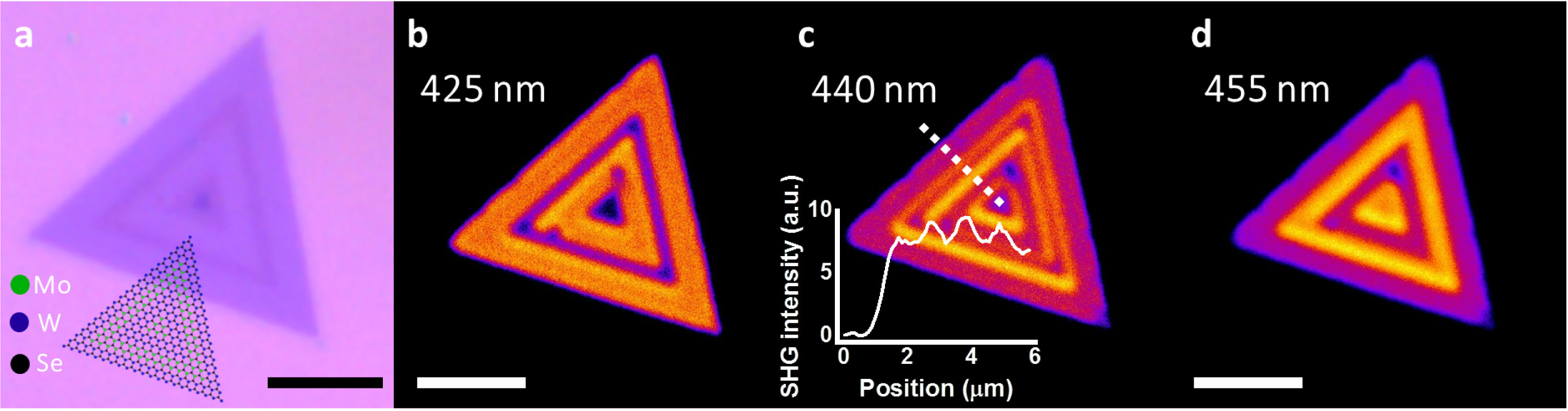} 
 \caption{{\small {\bf Optical and Second Harmonic imaging of a MoSe$_{2}$-WSe$_{2}$ based monolayer lateral heterostructure}. {\bf a} Optical image and schematic atomic structure of a three-junction MoSe$_{2}$-WSe$_{2}$ based 1L-LH. The 1L-MoSe$_{2}$ domain have a darker contrast with respect to the 1L-WSe$_{2}$  domain. From the center to the edge the regions are based on MoSe$_{2}$-WSe$_{2}$-MoSe$_{2}$-WSe$_{2}$. {\bf b-d} SH intensity images of the same 1L-LH sample collected at three different emission wavelengths. It can be noticed a greater SH intensity in the 1L-WSe$_{2}$ domain at 425 nm ({\bf b}), approximately the same SH intensity in both domains at 440 nm ({\bf c}) and a greater SH intensity in the 1L-MoSe$_{2}$ domain at 455 nm ({\bf d}). An enhancement of the SH intensity emerges at the heterojunctions when the SH intensities of both domains are similar. It is clearly noticed in the intensity profile of the dashed line shown in the inset of ({\bf c}). Scale bar: 5~$\mu$m.}}
 \label{fig1}
 \end{figure}

 Figure \ref{fig1}a shows an optical image of a three-junction 1L-LH composed by MoSe$_{2}$ and WSe$_{2}$ regions, and its schematic atomic structure. From the center to the edge the regions are based on MoSe$_{2}$-WSe$_{2}$-MoSe$_{2}$-WSe$_{2}$, where the 1L-MoSe$_{2}$ regions have a darker contrast with respect to the 1L-WSe$_{2}$ regions. As shown by Sahoo \textit{et al.}\cite{Sahoo2018}, each region consists of pure 1L-MoSe$_{2}$ or 1L-WSe$_{2}$ domains. In the junctions of the materials there is a chemical transition that displays different behaviors for each type of interface. The 1L-WSe$_{2}$ $\rightarrow$ 1L-MoSe$_{2}$ interface is atomically sharp, presenting an average width of 1 nm (4 atomic columns). The 1L-MoSe$_{2}$ $\rightarrow$ 1L-WSe$_{2}$ interfaces, in its turn, have a smoother chemical transition with some degree of alloy formation, with an average width of 6 nm (21 atomic columns) \cite{Sahoo2018}. 
 
 Figures \ref{fig1}b-d show the SH intensity images of the same 1L-LH in Figure \ref{fig1}a for three different emission wavelengths: 425, 440, and 455 nm, respectively.  Comparing these images, we can observe that the relative SH intensity of the materials change with the wavelength: 1L-WSe$_{2}$ has a greater SH intensity at 425 nm (Figure \ref{fig1}b), both materials have approximately the same SH intensities at 440 nm (Figure \ref{fig1}c), and 1L-MoSe$_{2}$ has a greater SH intensity at 455 nm (Figure \ref{fig1}d). It is important to note the appearance of a SH enhancement at the materials interface, as can be clearly seen in Figure \ref{fig1}c, when the SH intensities of both materials are similar. The SH intensity profile at the inset of Figure \ref{fig1}c shows that SH intensity is enhanced by approximately 23\% at the atomically sharp interfaces between the TMDs-domains.

\begin{figure}[!htb]
 \centering
 \includegraphics[scale=0.5]{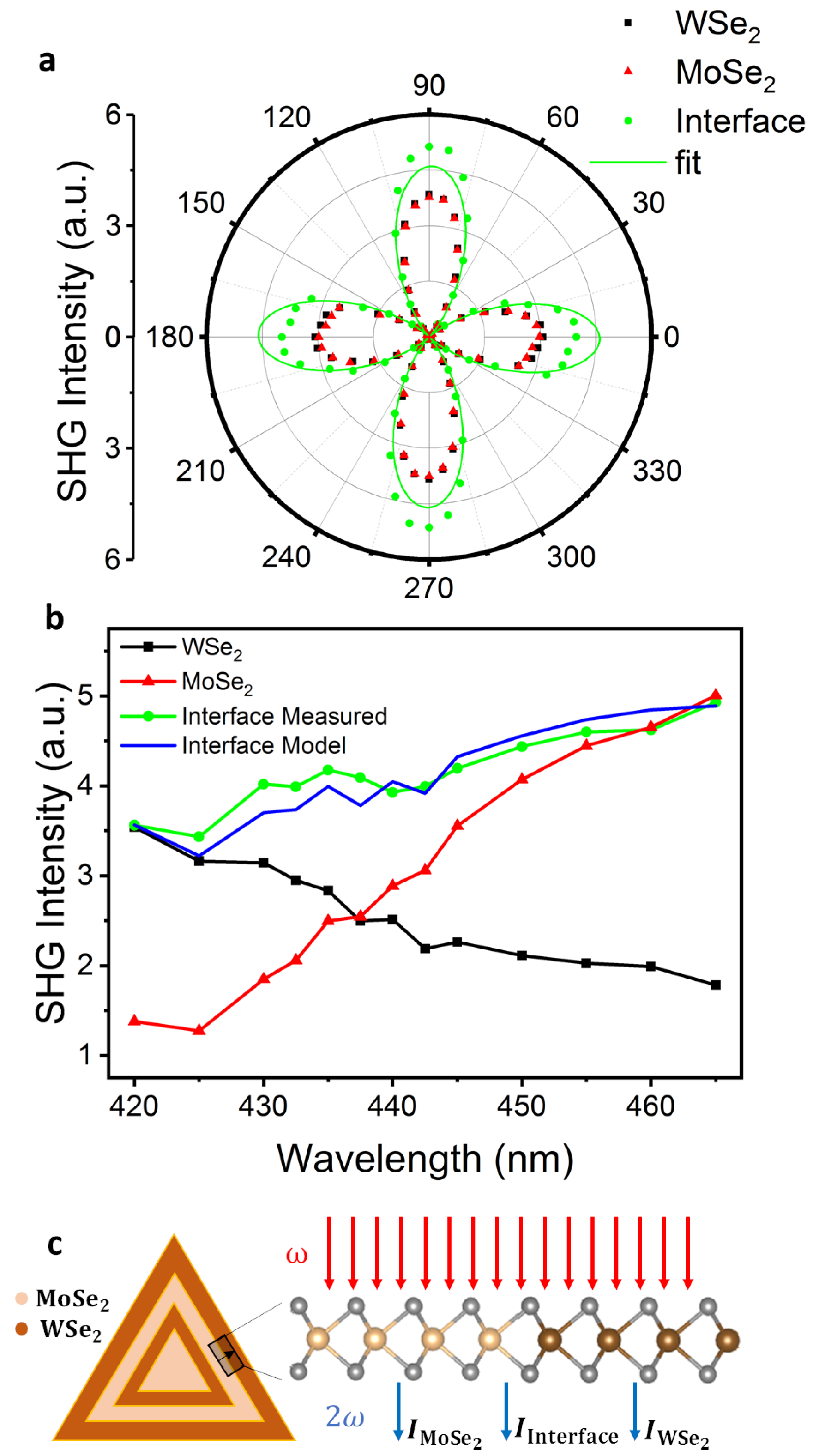} 
 \caption{{\small {\bf Second Harmonic polarization and wavelength dependent measurements in the MoSe$_{2}$-WSe$_{2}$ based monolayer lateral heterostructure}. {\bf a} Polarization-resolved SH measurement from the 1L-MoSe$_{2}$ and 1L-WSe$_{2}$ domains and from their interface for a 440 nm SH emission wavelength. The interface displays a greater SH intensity while both materials domains show similar SH intensities. Furthermore, the overlap of the four petals pattern for these three regions shows that they all have the same crystal orientation.  {\bf b} SH intensity as a function of the emission wavelength for the three regions cited above. It is noticeable the greater SH intensity at the interface where the SH intensities of both materials are similar (425-455 nm). {\bf c} A schematic illustration of the SHG at the 1L-LH: a laser beam with frequency $\omega$ excites the sample that generate a SH emission with frequency $2\omega$.}}
 \label{fig2}
 \end{figure}

 As shown by Yin \textit{et al.} \cite{Yin2014}, the crystal orientations of adjacent domains play an important role in the behavior of the SH signal at their interface. Therefore, to obtain better insights of the SH emission of our LHs, we performed a polarization-resolved SH measurement, that is widely used to ascertain the crystal orientation of 2D materials \cite{Li2013,Malard2013,Kumar2013}. For this measurement, we placed a rotable half-wave plate (600-1200 nm) before the objective to control the incident laser polarization at the sample and a fixed analyzer in front of the photomultiplier tube (PMT) in the horizontal direction. Figure \ref{fig2}a shows the polarization-resolved measurement of the same 1L-LH from Figure \ref{fig1} for a 440 nm SH emission wavelength. The measurements reveal the SH four-petals intensity pattern as expected from the calculations presented in Section S2 in the SI (the SH intensities were extracted as shown in SI Section S3). We can notice that all petals from the 1L-WSe$_{2}$ and 1L-MoSe$_{2}$ domains and from their interface overlap. It implies that they have parallel crystal orientations. It is important to note that $180^{o}$ crystal rotation would also give the same result for the SH polarization dependence. However, we do not observe the destructive interference between the domains as previous works \cite{Yin2014,bruno2020}. Moreover, it has been shown that the 1L-LHs domains have the same crystal orientations \cite{Gong2014,Huang2014,Gong2015,Duan2014,Zhang2015,Li2015,Sahoo2018}.

 To gain further information about the observed enhancement in the SH emission at the interfaces, we performed a set of measurements in the same 1L-LH with varying the pump wavelength. Since the SHG images are limited by optical resolution of 500 nm and the interfaces have a few nm width, we have deconvoluted the SHG images with the point spread function (PSF) of our system as presented in the Section S3 in the SI to improve the spatial resolution. Then, the mean values of the SH intensity at both materials domains (1L-WSe$_{2}$ and 1L-MoSe$_{2}$) and at their interface from the deconvoluted images were extracted (see Section S3 in the SI) for all measured wavelengths and presented in Figure \ref{fig2}b. We can observe that for wavelengths below 435 nm the SH intensity from 1L-WSe$_{2}$ domain predominates, whereas for wavelengths above 435 nm the 1L-MoSe$_{2}$ region has a greater SH intensity. The reason for this intensity behavior is due to the resonance of the two-photon process with the C exciton energy. The absorption maximum at the C exciton for 1L-WSe$_{2}$ and 1L-MoSe$_{2}$ are, respectively, 427 and 477 nm (see Section S4 in the SI). Thereby, each material is expected to have a greater SH intensity around its absorption peak wavelength. Besides that, at 435 nm, the absorption cross section for both materials are similar, giving rise to the same SH intensity as observed in Figures \ref{fig1}c and \ref{fig2}b. Moreover, Figure \ref{fig2}b reveals that the SH intensity enhancement at the interfaces can be clearly observed for SH wavelengths between 425 and 455 nm, having a higher intensity compared to the SH emission from the domains. This enhancement becomes stronger around 435 nm, where the SH intensities of both materials are similar, and disappear below 425 and above 455 nm, where there is a substantial difference between the SH intensities of the materials.

 To understand the observed SH enhancement, Figure \ref{fig2}c shows a schematic illustration of the SHG at the 1L-LH. The pump laser with frequency $\omega$ excites the sample that emits a SH signal with frequency 2$\omega$. When the laser beam is centered at the interface, the SH emission has contributions from both materials. As shown by Hsu \textit{et al.} \cite{Hsu2014}, the intensity of the SHG at the interface can be modeled as a coherent superposition of each material signal. Therefore, the electric field of the interface region $\Vec{E}_{int}(2\omega)$ is the superposition of the electric field of both materials: $\Vec{E}_{int}(2\omega) = \dfrac{\Vec{E}_{Mo}(2\omega)+\Vec{E}_{W}(2\omega)}{\sqrt{2}}$. This $(1/\sqrt{2})$ factor arises because each material contributes with half of its total emission intensity, that is proportional to the square of the absolute value of the electric field ($I \propto |\Vec{E}|^{2}$). Hence, the total SH intensity at the interface is given by:

 \begin{equation} \label{eq1}
     I_{int} = \frac{I_{Mo}}{2} + \frac{I_{W}}{2} + \gamma \sqrt{I_{Mo}I_{We}},
 \end{equation}

 \noindent in which $I_{Mo}$ and $I_{W}$ are the SH intensities from 1L-MoSe$_{2}$ and 1L-WSe$_{2}$ domains, respectively, and $I_{int}$ is the SH intensity at the interface. The $\gamma$ factor is introduced here to account for the phase difference between the two materials. This phase difference can be due to different crystallographic orientations \cite{Yin2014} or due to a intrinsic phase difference of different $\chi^{(2)}$ of the materials as shown by Kim {\it et al.}\cite{Kim2020}. Since both materials have the same crystallographic orientations \cite{Sahoo2018}, a deviation from $\gamma=1$ should be due to a phase difference coming from the distinct $\chi^{(2)}$ of both materials. In order to check if such simple model explains our findings, we calculated from Equation \ref{eq1} the expected SH interface intensity as a function of wavelength as shown in Figure \ref{fig2}b. The behavior of the interface SH intensity calculated by the interference model is in a reasonable agreement with the measured interface SH intensity for $\gamma = 0.5$. It is worth to comment that although we verify that the phase difference of SH fields of the two materials is important to explain our data, the exact value of this phase difference is not reliable due to limited spatial resolution of our experimental setup and small fluctuations in intensity for different interfaces as shown in Figure \ref{fig1}c.

\begin{figure}[!htb]
 \centering
 \includegraphics[scale=0.55]{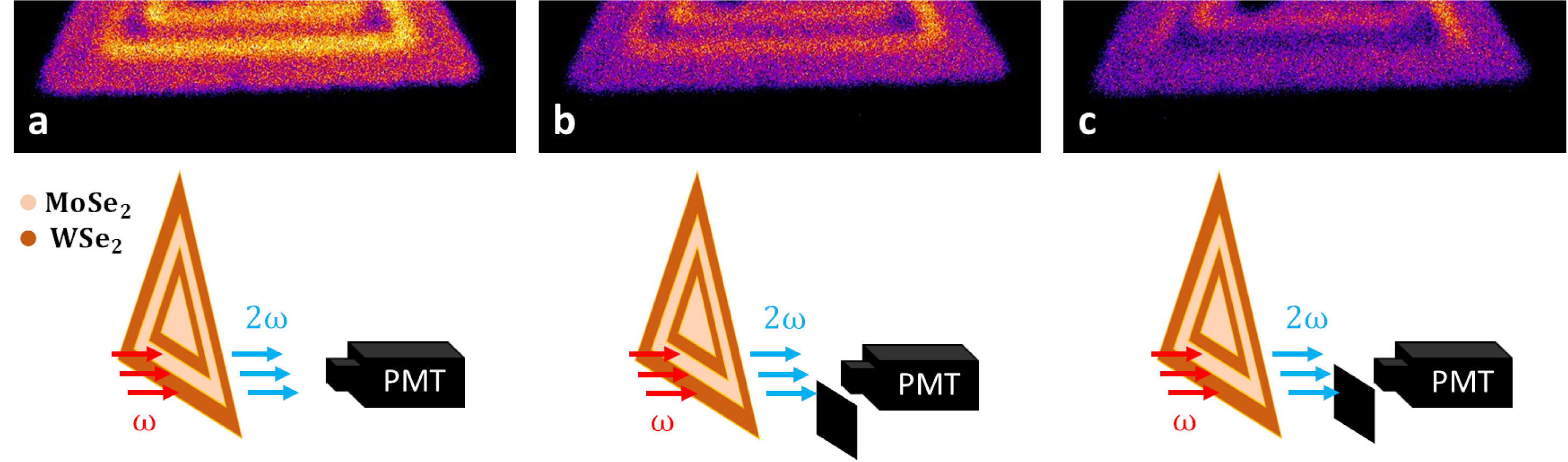} 
 \caption{\small {\bf Interference pattern evidence in the MoSe$_{2}$-WSe$_{2}$ based monolayer lateral heterostructure}. {\bf a-c} Schematic illustrations of a gradually blocked SH emission measurements and their respective SH images. In the absence of the blocking object ({\bf a}) or with it barely blocking part of the 1L-WSe$_{2}$ domain SH emission ({\bf b}), the SH enhancement at the interface is observed. Once there is a substantial blocking of the 1L-WSe$_{2}$ domain SH emission ({\bf c}), the interference pattern in the adjacent interface vanish. All SH images were collected at a 440 nm emission wavelength, where there is a clear SH intensity enhancement at the interfaces.}
 \label{fig3}
 \end{figure}

 To have a further experimental check that interference is the origin of the SH enhancement at the interfaces, we performed a set of SH measurements for a 440 nm emission wavelength by blocking the emitted light from one of the 1L-TMD domains as shown in Figure \ref{fig3}. When the SH is generated at the LH interface, the SHG beam has contributions from both domains that interfere with each other at the PMT detector. Therefore, it is expected that if we block the SH emission from one of the domains, the interference should disappear. To prove that, we have placed a movable sharp and straight object parallel to one of the MoSe$_{2}$-WSe$_{2}$ 1L-LH interfaces before the PMT detector as shown schematically at the bottom of Figure \ref{fig3}. Figure \ref{fig3}a shows the SH image with a clear SH enhancement at the interface without blocking the SH emissions. When the SH emission from one of the 1L-WSe$_{2}$ domains starts to be blocked, the SH intensity begins to decrease as shown in Figure \ref{fig3}b. If the blocking of the SH emission at the 1L-WSe$_{2}$ domain is increased, the SH enhancement at the adjacent interface is not visible anymore, as revealed in Figure \ref{fig3}c, confirming that this SHG enhancement at the interface comes from interference.

 In order to verify if the SHG can be used to probe other types of heterojunction interfaces, we also performed measurements in other LHs based on different TMDs (MoS$_{2}$ and WS$_{2}$) and for different number of layers. First, we performed SH measurements in a MoSe$_{2}$ and WSe$_{2}$ based bilayer (2L) LH. However, we did not observed any SH signal in our measurements, which led us to infer that this MoSe$_{2}$ and WSe$_{2}$ based 2L-LH have AB stacking that displays inversion symmetry and therefore does not generate SH emission. 
 
\begin{figure}[!htb]
 \centering
 \includegraphics[scale=0.5]{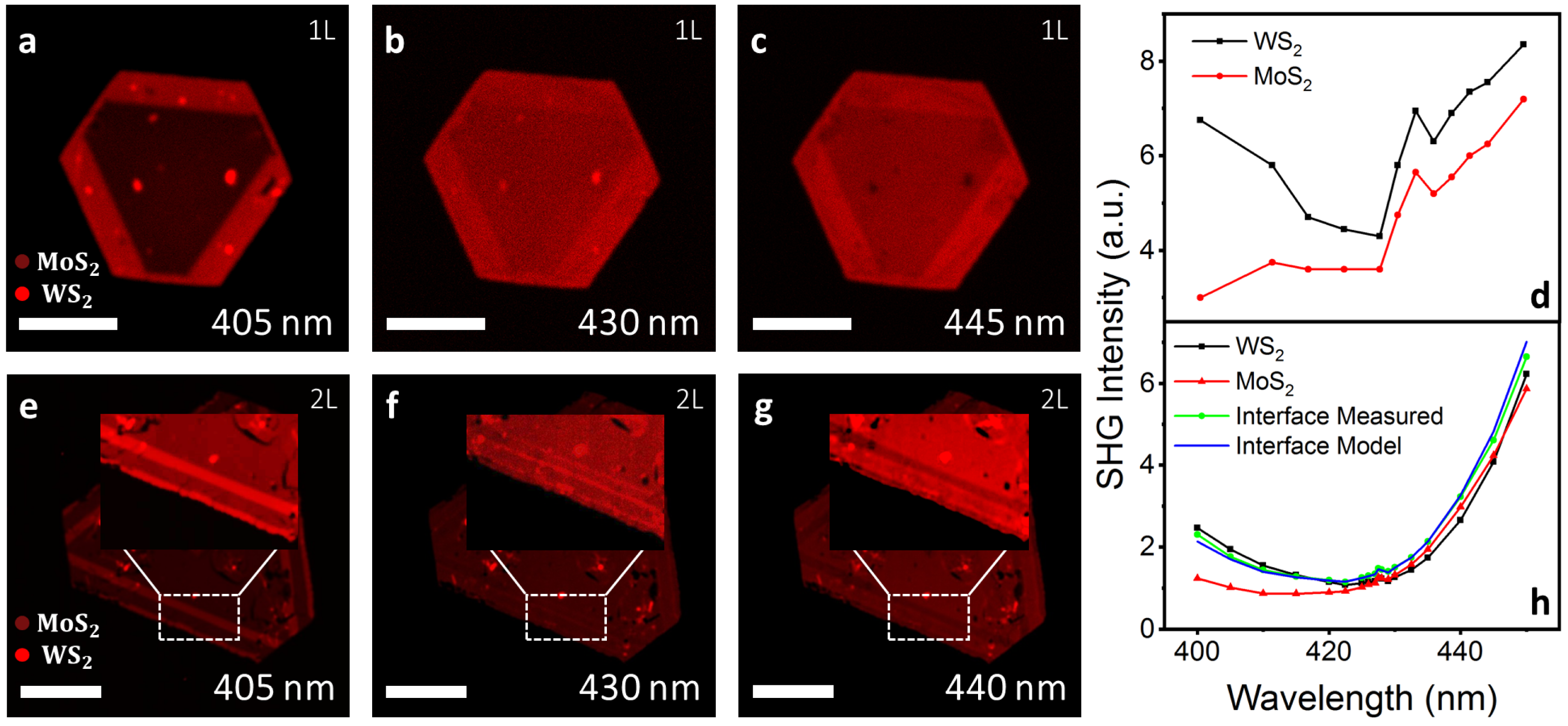} 
 \caption{{\small {\bf Second Harmonic intensity images and wavelength dependent measurements of MoS$_{2}$-WS$_{2}$ based monolayer and bilayer lateral heterostructures}. {\bf a-c} SH intensity images of the MoS$_{2}$-WS$_{2}$ based 1L-LH for three different emission wavelengths: 405 nm ({\bf a}), 430 nm ({\bf b}), and 445 nm ({\bf c}). The 1L-MoS$_{2}$ domain is in the center of the LH while the 1L-WS$_{2}$ domain is at the edge. {\bf d} SH intensity of the 1L-LH as a function of the emission wavelength for both materials. We can notice that the 1L-WS$_{2}$ SH intensity is greater for all the wavelengths measured, implying in no enhancement of the interface SH signal. {\bf e-g} SH intensity images of the MoS$_{2}$-WS$_{2}$ based 2L-LH for three different emission wavelengths: 405 nm ({\bf e}), 430 nm ({\bf f}), and 440 nm ({\bf g}). From the center to the edge the regions are composed by MoS$_{2}$-WS$_{2}$-MoS$_{2}$-WS$_{2}$. The insets show more clearly the heterojunctions in the region indicated by the dashed lines. {\bf h} SH intensity of the 2L-LH as a function of the emission wavelength for both materials, their interface, and the interference model using $\gamma = 0.16$. For the 2L-LH there is an evident enhancement of the interface SH intensity around 430 nm, where the SH intensities of both materials are similar. Scale bar: 10~$\mu m$ ({\bf a-c}) and 20~$\mu m$ ({\bf e-g}).}}
 \label{fig4}
 \end{figure}

 On the other hand, Figure \ref{fig4} shows the measurements of MoS$_{2}$ and WS$_{2}$ based 1L- and 2L-LHs with finite SH signal. Figures \ref{fig4}a-c show SH intensity images of a MoS$_{2}$ and WS$_{2}$ based single-junction 1L-LH. The 1L-MoS$_{2}$ domain is in the center of the LH while the 1L-WS$_{2}$ domain is at the edge. Although the intensity ratio between both materials change with the emission wavelength, the 1L-WS$_{2}$ SH intensity is always greater and no enhancement of the SH emission was observed at the interfaces. Figures \ref{fig4}e-g show SH intensity images of a MoS$_{2}$ and WS$_{2}$ based three-junction 2L-LH with AA stacking. From the center to the edge the regions are composed by MoS$_{2}$-WS$_{2}$-MoS$_{2}$-WS$_{2}$. For this 2L-LH there is a clear change in the predominant material with stronger SH emission as a function of wavelength. 2L-WS$_{2}$ has a larger SH intensity at 405 nm (Figure \ref{fig4}e), both materials have approximately the same SH intensities at 430 nm (Figure \ref{fig4}f) and 2L-MoS$_{2}$ has larger SH intensity at 440 nm (Figure \ref{fig4}g). It is possible to observe in the inset of Figure \ref{fig4}f that there is an evident SH enhancement at the interfaces when the SH intensities at both domains are similar.

 As done for the selenide TMDs based 1L-LH, we performed a set of SH measurements varying the pump laser wavelength for both 1L- and 2L-LHs based on the sulfide TMDs. Figure \ref{fig4}d shows the 1L-MoS$_{2}$ and 1L-WS$_{2}$ domains SH spectra, where it is possible to observe that the WS$_{2}$ domain have a greater SH intensity over the whole wavelength interval measured. This is different from what is observed at the MoSe$_{2}$-WSe$_{2}$ based 1L-LH SH spectra (Fig. \ref{fig2}b), where there is an inversion in the domain with predominant SH emission. Therefore, although the interference model (Equation \ref{eq1}) remains valid for the MoS$_{2}$-WS$_{2}$ based 1L-LH, the absence of a spectral region where the SH intensities of both domains are similar explains why there is no visible enhancement in the SH emission at the interfaces, since the interference is not completely constructive ($\gamma<1$). It is worth to note that the lack of an intersecting point between the 1L-MoS$_{2}$ and 1L-WS$_{2}$ SH spectra is also observed in their absorption spectra for this same wavelength range, as shown in Section S4 in the SI.
 
 Conversely, the MoS$_{2}$-WS$_{2}$ based 2L-LH presented a clear SH enhancement at their interfaces (Figure \ref{fig4}f). Hence, we have deconvoluted their SH images with the PSF of the system to improve the heterojunctions resolution. The intensities of both materials and of the interface for the wavelength dependent measurements were extract in the same way as before (see Section S3 in the SI) and their SH spectra are shown in Figure \ref{fig4}h. We have also calculated the expected value of the interface intensity for this MoS$_{2}$-WS$_{2}$ based 2L-LH using the interference model (Equation \ref{eq1}), and it is also shown in Figure \ref{fig4}h. We can notice that the interference model is in agreement with the measured SH intensities values at the interface by using $\gamma = 0.16$. Polarization-resolved measurements were also performed in the MoS$_{2}$-WS$_{2}$ based 2L-LH to ascertain the domains and interface crystal orientations. As presented in Section S5 in the SI, the regions have all the same crystal orientation.

 In summary, we have observed a constructive interference emerging from the coherent superposition of the SH signal from each material at their atomically sharp interfaces for different TMD LHs. Performing polarization-resolved and energy dependence measurements we were able to model this interference pattern and notice a phase difference between materials in the LHs due to their different $\chi^{(2)}$. For the MoSe$_{2}$-WSe$_{2}$ based 1L-LH and the  MoS$_{2}$-WS$_{2}$ based 2L-LH, we have observed an enhancement of the SH emission at the interface with respect to the SH emission from both materials for a certain pump wavelength range. This enhancement only occurs when the SH emission from each material is similar in intensity, leading to a highlighted constructive interference. However, even with no enhanced SH signal, the constructive interference is also supposed to happen at the interfaces, as in the MoS$_{2}$-WS$_{2}$ based 1L-LH. Therefore, our model enable us to infer that any LH would present a constructive interference at their interfaces if their domains have the same crystal orientations, but it might be not visible as in the case of MoS$_{2}$-WS$_{2}$ based 1L-LH. Moreover, our results show that SHG, a non-destructive optical technique, can also be used to image sharp interfaces, of nanometric widths, in LHs based on 2D materials.

\section{Methods}

 \subsection{Sample Preparation}

  Lateral heterostructures used for this study were synthesized directly via water-assisted one-pot chemical vapor deposition (CVD) method at atmospheric pressure, as per our previous report \cite{Sahoo2018, Sahoo2019, Berweger2020}. Bulk powders of either MoS$_{2}$ and WS$_{2}$ or MoSe$_{2}$ and WSe$_{2}$ were used directly as solid precursors for the growth of either MoS$_{2}$-WS$_{2}$ or MoSe$_{2}$-WSe$_{2}$ heterostructures, respectively. In brief, for the growth of MoX$_{2}$-WX$_{2}$ [where X=S or Se] LHs, a high pure ceramic boat containing a mixture of MoX$_{2}$ and WX$_{2}$ powders was placed within a two-zone quartz tube furnace (diameter 1”).  Pre-cleaned SiO$_{2}$/Si substrates, with acetone, isopropanol and deionized water, were used for the direct deposition of vapor phase precursors followed by the growth of heterostructures. During the growth, the solid sources were processed at 1060 $^{o}$C, whereas the growth substrates were located downstream at 6-10 cm away and at temperature of 800-750 $^{o}$C. Initially, the temperature of the solid source zone was slowly elevated to 1060 $^{o}$C in 45 min with a constant 200 SCCM N$_{2}$ flow, while both, substrates and sources, were kept outside of the hot zone. When the temperature of the hot zone reached above 1000 $^{o}$C, the quartz tube containing solid precursor and the substrates were placed to respective processing temperature carefully, by sliding the tube into the furnace, and instantaneously water vapor was introduced by changing the path of N$_{2}$ flow through a bubbler containing 2 ml of deionized water at room temperature. During the presence of water vapor, only MoX$_{2}$ domains were formed on the substrates. To switch the synthesis condition from Mo- to W-rich compounds for the growth of LHs, the N$_{2}$ + H$_{2}$O flow was suddenly substituted by a mixture of Ar + 5$\%$ H$_{2}$ (200 sccm) which promoted the growth of WX$_{2}$ domains at the edges of pre-existing MoX$_{2}$ domains. The size and width of individual domains within a lateral heterostructure could be controlled by changing the deposition rate, whereas the number of junctions were controlled only sequentially changing the carrier gas between N$_{2}$ + H$_{2}$O and Ar + 5$\%$ H$_{2}$. Once the desired heterostructure sequences were completed, the growth process was abruptly terminated by sliding the quartz tube further to a cooler zone under the flow of Ar + H$_{2}$ (5$\%$), until it cooled down to room temperature.

 \subsection{Nonlinear Optical Spectroscopy}

  SHG experiments were done by using an optical parametric oscillator system (APE picoEmerald) tunable from 750 to 950 nm, with 7 picosecond pulse width, and 80 MHz repetition rate. SH imaging was performed by scanning the laser with a set of galvanometric mirrors (LaVison BioTec) in a Nikon microscope. The laser beam was focused on the sample by a 40x objective with numerical aperture N.A. = 0.95. The backscattered SH signal was collected by the same objective, reflected by a dichroic mirror that reflects bellow 690 nm and then directed to a PMT. The SH images were made by an image acquisition software (LaVision BioTec Imspector Pro). The laser power at the sample was kept constant and bellow 5 mW for all measurements in the same heterostructure. For the polarization-resolved measurements, the SH intensity images were made with a step of $3^{o}$ in the half-wave plate. The SH intensities were analyzed using the ImageJ software.
    
 \subsection{Raman Spectroscopy}

  Raman spectra were obtained in a confocal Raman spectrometer (WITec Alpha 300R) using a 532 nm laser line and a $100\times$ objective lens. The laser power was kept at 0.5 mW during all measurements.

\begin{acknowledgement}
The Brazilian authors acknowledge financial support from CNPq, Brazilian Institute of Science and Technology (INCT) in Carbon Nanomaterials, CAPES, FAPEMIG and FINEP. We also acknowledge the LCPNano multiuser laboratory from the Physics Department at UFMG for Raman spectroscopy characterization. A.R.C. acknowledges the FAPESP fellowship (Grant No.  2020/04374-6). P.K.S. acknowledges the ISIRD start-up grant from IIT Kharagpur. 

\end{acknowledgement}

\begin{suppinfo}

Raman spectra, optical images and Raman peak positions of MoSe$_{2}$-WSe$_{2}$ based 1L-LH, MoS$_{2}$-WS$_{2}$ based 1L-LH, and MoS$_{2}$-WS$_{2}$ based 2L-LH. SH polarization calculation. SH raw image, PSF, and SH deconvoluted image of MoSe$_{2}$-WSe$_{2}$ based 1L-LH. SH intensity analysis. MoSe$_{2}$, WSe$_{2}$, MoS$_{2}$, and WS$_{2}$ absorption spectra at C exciton. Polarization-resolved SH measurement of MoS$_{2}$-WS$_{2}$ based 2L-LH.

\end{suppinfo}

\providecommand{\latin}[1]{#1}
\makeatletter
\providecommand{\doi}
  {\begingroup\let\do\@makeother\dospecials
  \catcode`\{=1 \catcode`\}=2 \doi@aux}
\providecommand{\doi@aux}[1]{\endgroup\texttt{#1}}
\makeatother
\providecommand*\mcitethebibliography{\thebibliography}
\csname @ifundefined\endcsname{endmcitethebibliography}
  {\let\endmcitethebibliography\endthebibliography}{}

\end{document}